**Article title**
Recovering the polyhedral geometry of fragments

**Authors**
*János Török* * (1,2), *Gábor Domokos*(1,3)

**Affiliations**
(1) HUN-REN--BME Morphodynamics Research Group, Budapest University of Technology and Economics, Műegyetem rakpart 1-3, Budapest 1111, Hungary
(2) Department of Theoretical Physics, Institute of Physics, Budapest University of Technology and Economics, Műegyetem rakpart 1-3, Budapest 1111, Hungary
(3) Department ofMorphology and Geometric Modeling, Budapest University of Technology and Economics, Műegyetem rakpart 1-3, Budapest 1111, Hungary

**Corresponding author's email address and Twitter handle**
torok.janos@ttk.bme.hu

**Abstract**
Not only is the geometry of rock fragments often well approximated by *ideal* convex polyhedra having few faces and vertices, but these numbers carry vital geophysical information on the fragmentation process. Despite their significance, the identification of the number of faces and vertices of the ideal polyhedron has so far been carried out only through visual inspection. Here, we present an algorithm capable of performing this task in a reliable manner. The input of our algorithm is a 3D scan of the fragment which is essentially a triangulated polyhedron, which however has often large number of faces and vertices. Our algorithm performs a systematic simplification using the following steps:
- Gaussian smoothing is performed on the spherical histogram of the 3D scans faces to identify the most important face orientations.
- Planes carrying the faces of the ideal polyhedron are identified as maxima of the smoothed histogram
- Polygon is reconstructed using the identified planes
- Small faces are removed in a systematic manner

We present two versions of the algorithm that we benchmarked the algorithm against a dataset of human measurements on 132 fragments. Beyond identifying the ideal polyhedral approximation for fragments, our method is also capable of tracing backward the shape evolution of rounded pebbles to their origins.

**Graphical abstract**

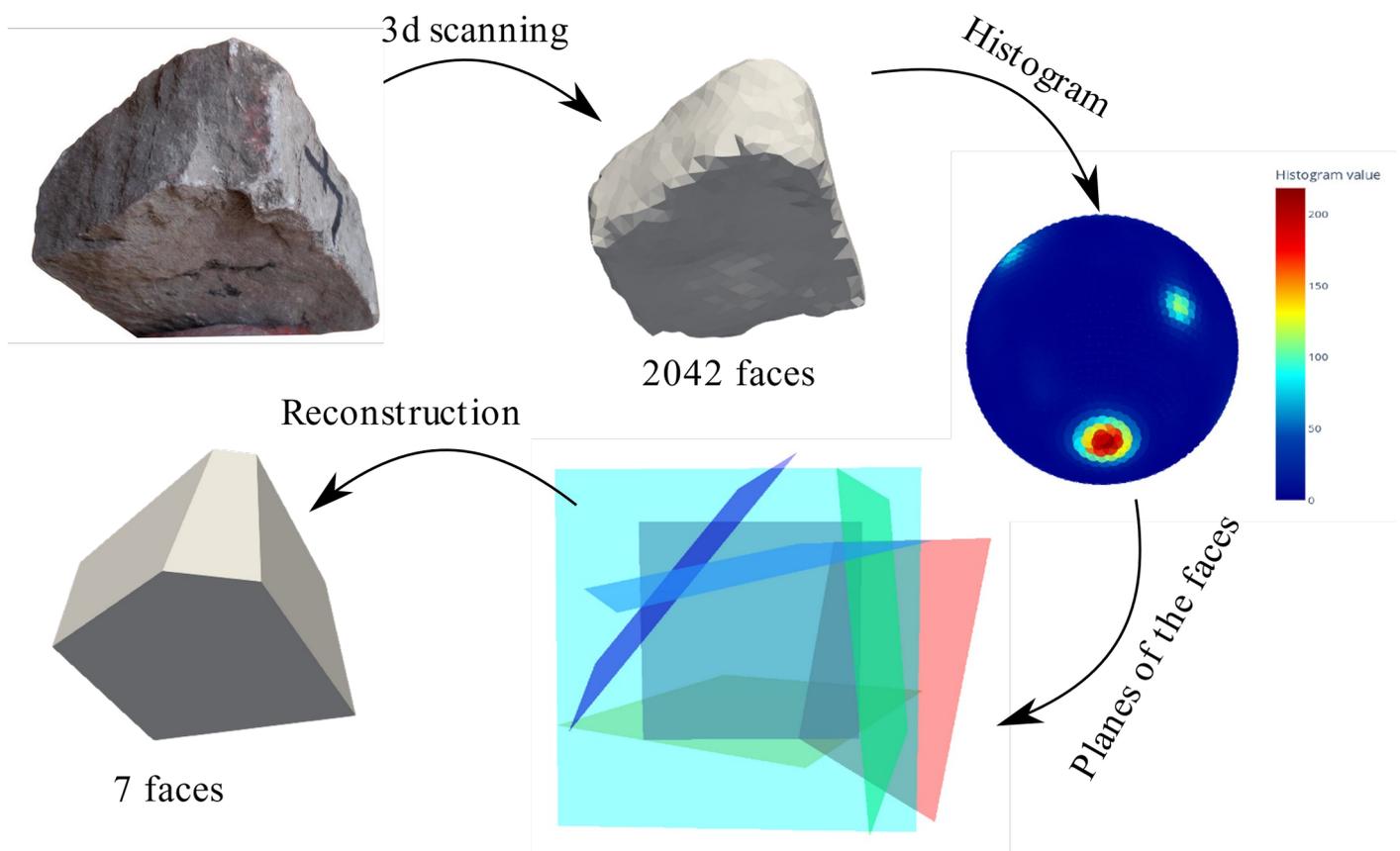

**Specifications table**

| Subject area | Earth and Planetary Sciences |
|---|---|
| More specific subject area | Fragmentation |
| Name of your method | polyhedronizer |
| Name and reference of original method | - |
| Resource availability | Source code: https://github.com/torokjanos-bme/polyhedronizer, Validation data: https://osf.io/h2ezc/ |

**Background**

As observed in [Domokos 2020], the fragmentation of homogeneous, isotropic rock under a uniform stress field (to which we jointly refer as *ideal* conditions) produces convex polyhedra as fragments and the combinatorial properties (e.g. numbers of faces and vertices) of these polyhedra carry vital (and, sometimes, the only) information about the provenance of the fragments: combinatorial numbers can tell the stress field which produced the fragments. For example, hydrostatic stress splits the rock into a Voronoi-type tessellation with polyhedral fragments with respective average numbers for faces and vertices $(\bar{f}, \bar{v})=(15.51, 27.07)$. On the other hand, multiple, successive shear events result in $(\bar{f}, \bar{v})=(6,8)$. This illustrates that the identification of these numbers is of key importance if we want to use the geometry of the fragments to recover the geophysical process which has produced them. Nevertheless, the identification of the ideal convex polyhedron (carrying these key numbers) is far from trivial, as rock material is neither perfectly homogeneous, nor isotropic, nor is the stress field causing the fragmentation perfectly uniform: in brief, conditions are not ideal nor are the resulting fragments. Our current methods article addresses this problem.

While the physical fragments we observe are visually close to a convex polyhedron, nevertheless, due to the aforementioned material and physical imperfections, the fragment shapes carry many indentations and irregularities. Recovery of the ideal, convex polyhedral shape can be done by manual count using visual inspection. While such estimates produce good approximations, they however depend on the person and thus they are hard to verify and to compare. Since 3D imaging technology is readily available [Fehér 2023], here we present an algorithm and its implementation which recovers the ideal, underlying convex polyhedron in a transparent manner. Our method assigns an ideal convex polyhedron to any scanned particle and also provides error estimates to gauge the reliability of the output. The method has two essential control parameters which we tuned, using a large dataset from manual/visual measurements as benchmark. Since the configuration space of convex polyhedral is large (for a polyhedron with *E* edges this space has *d=E-1* dimensions), achieving good match with just two control parameters appears reassuring and suggests that the results are robust and reliable.

Often, scanned fragments are represented by a triangulated surface. In the first step, we construct the convex hull [Sherman 1955]. This has double purpose, first, it ensures that the direction of the surface elements of the object actually align with the surface which is not the case if the fragment have surface irregularities due to material grain structure, second, since we are interested in convex representation of the objects the indentations due to impurities are meaningless for our analysis. The main idea of our method is using statistics of the outer normal of this triangulated convex boundary.

We have validated our method on a dataset produced by manual measurements for [Domokos 2020] (see section Method validation). The dataset contains 132 fragments from the Hármashatárhegy mountain in Budapest, Hungary. The composition of the fragments is dolomite. The fragments were collected from weather induced fractured areas. All fragments were scanned in three dimensions and we compared the output of our algorithm with the manual/visual data obtained by an expert technician.

**Method details**

The main purpose of this method is to recover the original (or ideal) convex polyhedral form associated with a scanned object. As noted, the ideal convex polyhedron is directly linked to the generating stress field. We assume that the input is a scanned object *S* the surface and the point cloud is associated with a triangulated mesh of its convex hull *C(S)*. The main steps of the algorithm are the following:

1. We consider the outer normals of the triangles associated with *C(S)* and construct a spherical histogram *H(C(S))* based on these normals. We also identify a reference point $O$, the coordinates of which are the averages of the respective coordinates in our point cloud. The point $O$ is always inside the convex hull of the object.
2. We perform Gaussian smoothing [Simonoff 2012] with parameter $\sigma$ on the spherical histogram *H* to obtain the smoothed histogram $H(\sigma)$. Note that $H(\sigma)$ is a scalar function on the sphere.
3. We identify the local maxima of $H(\sigma)$. We denote the number of the local maxima by the integer-valued function $F(\sigma)$.
4. Based on $F(\sigma)$, we determine the number $F$ of the faces of the ideal convex polyhedron. This can be done in two alternative manners:
   (a) We identify $F$ as the length of the longest plateau of $F(\sigma)$ or
   (b) We give an a priori bound $F_{max}$ and we find the closest value $F \leq F_{max}$ of $F(\sigma)$.
5. By determining the number $F$ in this manner, we also determined orientation of the $F$ faces. Next we determine the outer normal associated with them.
6. We determine the distance $d_i (i=1,2,\ldots F)$ of each face from the reference point $O$.
7. We construct a hyperplane mosaic [Domokos et al 2020] $M_F$ consisting of the $F$ selected planes. Note that all cells of $M_F$ are convex, some are bounded, some are unbounded.
8. We determine all bounded cells of $M_F$. All bounded cells are convex polyhedra.
9. We pick the unique bounded cell which contains the reference point $O$. If there is no such cell, then the set of $F$ planes is not a real representation of a polyhedron, larger value of $F$ has to be chosen.
10. Post-processing: we assume that sufficiently small faces have been created by *chipping events* where the polyhedron was bisected by a plane. We assign unit volume to the polyhedron with $F$ faces and under this assumption, for each sufficiently small face we reconstruct the *other, missing* part of the polyhedron. We successively remove faces but only if the volume increases less than a given factor *dv*.

Our algorithm has two major control parameters. For version (b) we have $F_{max}$ and for both versions we have *dv*.

One important consequence of our algorithm is that the predicted polyhedra are always simple, namely that exactly three faces and three edges meet at every vertex. This is the result of the reconstruction process which identifies the planes of the sides and not the corners of the polyhedron. In [Domokos 2020] it was shown that this process always leads to a simple polyhedron. A simple polyhedron can be characterized by a single parameter e.g. the number $F$ of faces, since the number *V* of vertices can be calculated as *V=2F-4*. from the simplicity property of the polyhedron and Euler formula.

The difference between the versions (a) and (b) is twofold. First, for our data set algorithm (a) runs 44 minutes for determining the faces and 20 minutes to remove excess faces, (b) needed 52 minutes for the faces and 272 minutes to remove the unnecessary faces, which is roughly a factor 5 in running time. Second, the results differed in a way which is illustrated in Fig. 1, where the left figure was created with algorithm (a) the right one with (b). The original pebble has a wedge shape, and on the left hand side it has a rounded edge due to wear. The examiner did not identify this as a face, therefore his judgment on the values of $F$ matches with algorithm (a). On the other hand, the volume of model (a) is 35% larger than the original pebble, whereas for algorithm (b) it is only 20%. This difference is the result of whether the unrealistic wedge is chipped of or not by the algorithm. The most extreme volume difference was 660% in (a), vs. 30% in (b) for sample 84. We note that it is very likely that at the birth of the fragment the long wedge was indeed part of the fragment but it broke after the first few impacts the fragment had.

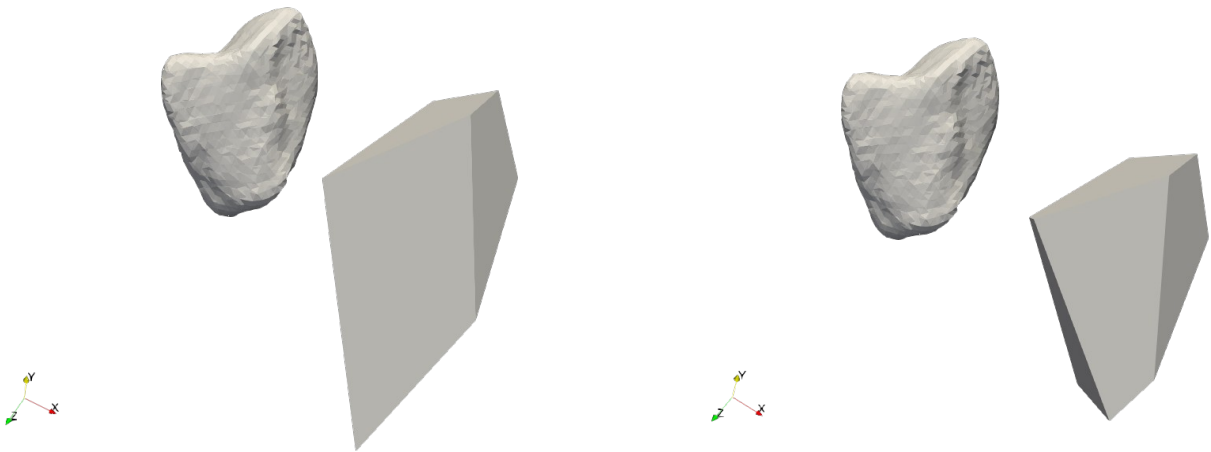

**Figure 1.** Left, the polyhedron from sample 70 with model (a), right with model (b). The difference is an extra side in algorithm (b) which cuts the wedge short to avoid excess volume. This was not identified as a real polygon side by the expert.

**In the following we detail the algorithm**

**0. Preprocessing**

Our code assumes the three-dimensional body to be described by polygons which are determined by their normals and corner points. Also it is assumed that the faces of the triangles are in general parallel to the actual surface they are on so working with convex hull is advisable. From now on we assume that we have a list of faces bounding the object, each parallel to the local shape and characterized by two quantities: the outer normal of the faces and the coordinates of the vertices. . Using the vertices, the surface area of the faces can be calculated.

1. **Spherical histogram**

The next step is to make a weighted histogram from the normals with the surface area as the weight. The most important problem here is to make reliable binning on the surface of the unit sphere. The Euler angles ( $\theta$ , $\phi$ ) are unsuitable for this purpose since they show singular behavior at the north and south pole, and it is very difficult to create equal sized bins. Therefore we used a special tessellation of the surface of the unit sphere which was proven to have 40% more uniformly distributed cells than any conventional polar tessellation, the Fibonacci lattice [González 2009]. This tessellation uses the angle of the golden ration: $\phi=1+\phi^{-1}=\frac{(1+\sqrt{5})}{2}$. Only odd number of points can be created which we denote as $P=2N+1$. The Euler angles at the meshpoints can be obtained by the following formula (assuming the periodicity of $\phi$).

$$\theta_i = \arccos\left(\frac{2i}{2N+1}\right), \tag{1}$$

$$\phi_i = 2\pi i \phi^{-1}, \tag{2}$$

where $i=1,2,\ldots P$..

This algorithm gives thus *P* bin points on the sphere and, according to [González 2009], the distribution of these points on the sphere is close to uniform. Figure 2. illustrates the positions of the grid points of this algorithm for $N=1000$.

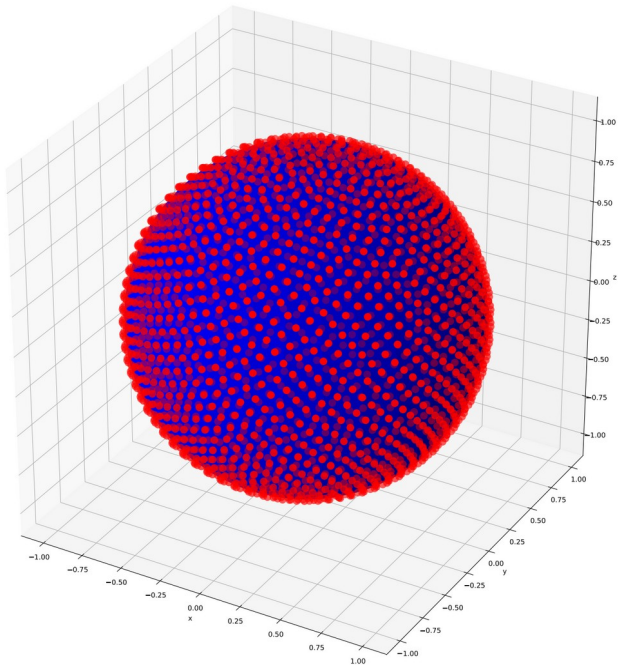

**Figure 2.** Positions of the Fibonacci lattice points on the unit sphere for $N=1000$.

The Fibonacci lattice points will serve as the bins of the histogram for the surface polygons. The procedure is the following: For the normal of each surface polygon we find the bin point which is closest to it (the dot product is maximal) and the value in the bin is increased by the surface area of the polygon (weighted histogram).

**2. Smoothing of the histogram**

The histogram is very noisy if *N* is large, on the other hand, decreasing *N* gives lower resolution which means that the sides of the reconstructed polyhedra will be very inaccurate in their orientation. Our solution to this problem is to use a smoothing function, or kernel [Simonoff 2012]. We chose a Gaussian kernel with the distance between two bin points defined as the shortest path on the surface of the unit sphere. The standard deviation $\sigma$ of the Gaussian kernel was varied in small steps in the range $\sigma \in [r/7, r]$, the lower value resulting in surface distance of a few bin points. Larger parameter range would not add reasonable solutions.

We have calculated the histogram for $N=500, 1000, 2000$ in method (a) and for $N=2000$ in method (b). In all cases we used we increased the standard deviation $\sigma$ for the Gaussian kernel in 50 equidistant steps from 1/7 to 1. The range of Gaussian kernel standard deviation $\sigma$ is not considered as a model parameter for two reasons: (a) While we consider the emerging polyhedral structure to be a function of the value of the $\sigma$, the final, optimal polyhedron will be selected based on extremal properties of this function (i.e. at the longest plateau) WE also mention that the range for $\sigma$ chosen by our algorithm covers the distance from grid level to radius scale, and therefore all possible scenarios are covered.

### 3. Finding local maxima

In the Fibonacci lattice we determine the distance $d_{nn}$ in such a way that all lattice points have in average 6 others within this distance. All points within $d_{nn}$ distance are considered to be neighbors. A bin point is considered a local minimum if all its neighbors contain smaller values.

### 4. Determining the number of faces

**Method (a):** The above histograms and local maximum analysis returns a value $F$ for the number of faces for all values of *N* and sigma. We observed that in most cases, if the $F$ was plotted against $\sigma$, the curve displayed a long plateau (see the left panel of Fig. 3.) We can observe that the number of bin points does not have a significant influence on the results, the plateau is robust and obvious. In spite of this we performed the computation for all three N values and identified $F$ as the value I which appeared most in all computations. In some cases this process does not lead to such an obvious plateau (see Fig. 3 right panel), in such cases Step 9 of the algorithm offers the solution. These cases correspond to more rounded objects, where the number of faces is not obvious, even for a human observers.

There were few cases where the faces created in this way did not result in a closed object (especially if one face is rounded and the neighboring faces intersect on the other side of the object unlike the ones in Fig. 1), in these cases we considered the second longest plateau, corresponding to a higher value of $F$.

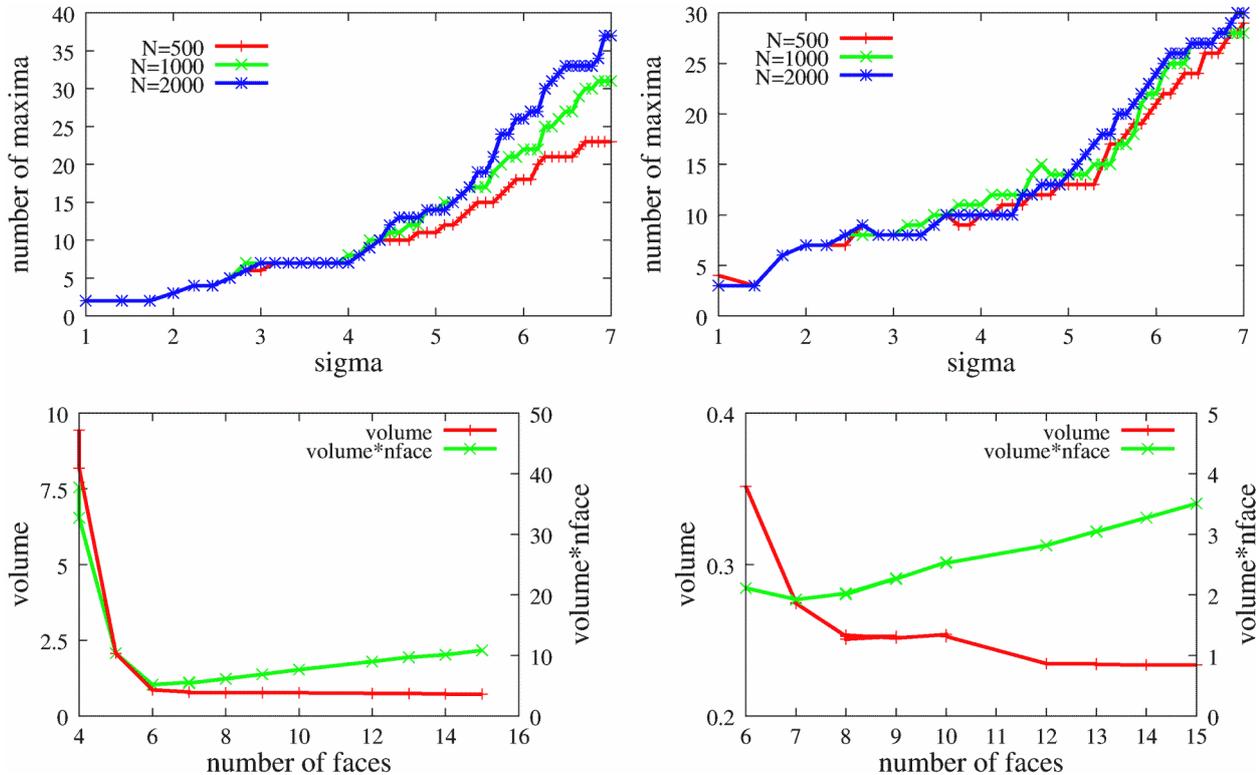

**Figure 3.** Top row: The number of local maxima in the histogram as function of the sigma parameter of the Gaussian kernel. The colors and symbols are for systems with different number of bin points are indicated in the legend in the top left corner of the plot.
Bottom row: Red curve with left y axes: The volume of the resulting polyhedron as function of the number of faces in arbitrary units. Green curve with the right y axes: The volume times the number of faces of the polyhedron as function of the number of faces.
Left: sample no. 9, right: no. 8.

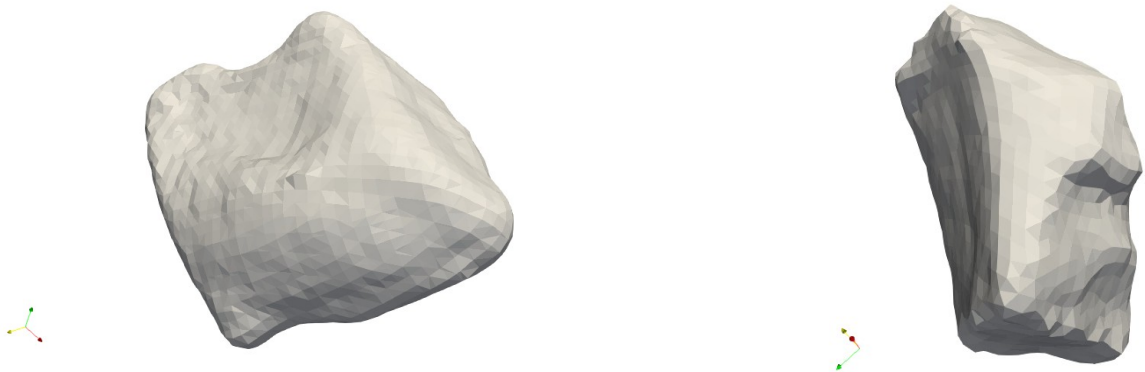

**Figure 4.** The pebbles 9 and 8. The sample on the left is a cuboid, the one a right hand side has a more complex structure.

**Method (b):** we chose the largest number of detected maxima under 16 for which we found a closed polyhedron.

**5. Determining the normals of the faces of the new polyhedron**

Once the number $F$ of faces has been is determined, we have identified the corresponding bin points on the $N=2000$ histogram. We identified the normals of the faces of the polyhedron by the vectors corresponding to the bin points.

**6. The distance of the faces from the centre of mass**

In order to determine the distance of a face from the center of mass of the original object, we have to measure the distance of the faces of the convex hull having normal close to the normal of the investigated face. In the specific examples we have considered two unit normals $n_1, n_2$ to be close if $n_1 n_2 \geq 0.95$ (but $n_1 n_2 \geq 0.99$ worked almost identically).
This step assigns subsets of faces of the original convex hull to each new polyhedral face. For each subset of complex hull faces we computed a weighted histogram, based on their distance to the origin, using the area of the faces as a weight function. We identified the distance by the position of the maximum in the histogram computed in the described manner, using B = 10 equal bins spanning between the largest and the smallest distance. The result is a set of $F$ distances which we denoted by $d_i = 1, 2, \ldots F$.

**7. The full 3d space is then subdivided by the computed planes, and all closed polyhedra are determined**

In this step we used the algorithm developed in [Domokos 2020] to determine the closed polyhedra. The input of this algorithm is a set of normals $n_j$ and the corresponding distances $d_i$ of the planes which contain the faces of the polyhedron. It is important that no two normals are parallel (although they can have opposite directions). The algorithm determines the relevant geometric objects in the following order:
  (a) Vertices
  (b) Edges
  (c) Faces
  (d) Polyhedra

(a) Vertices

Any set of three planes meet in one point. We solved $\frac{1}{6} s(s-1)(s-2)$ sets of linear equations and stored all points as possible vertices, together with the identifying indices of the corresponding three planes.

(b) Edges

The set of vertices which were obtained by the intersection of two given planes are collinear and thus they can be ordered using scalar product and two reference points. Once the points are ordered the successive pairs form possible edges. They are stored along with the indices of the two corresponding planes.

(c) Faces

Faces are a series of connected edges forming a closed convex polygon. We start with two adjacent edges that share a common vertex. These two edges determine the plane of the face. Additional edges are added to one end of this polygonal line while ensuring convexity and maintaining the condition that all edges are coplanar: The latter condition is guaranteed by appropriate selection from the list of available edges. On a planar section the edges of the emerging planar mosaic always form a vertex with shape "X", with exactly four edges that are pairwise aligned. In theory, selecting the edge that maintains convexity with the previous edges is straightforward (by checking the sign of the cross products). However, in computational applications, parallel lines do not exist in an exact sense, therefore we select the edge for which the cross product of the normals of the edges yields the smallest absolute value.

This algorithm has two possible stopping scenarios: (i) the next vertex of the newly added edge is the same as the first one of the first edge, then we made a loop, so we found a face, we store it with the plane index and the list of edges. (ii) we do not find any edge which we can connect to the last one in which case this is an open face and it is discarded.

(d) Polyhedra
Polyhedra created by intersecting planes are necessarily simple polyhedra, meaning that exactly three faces meet at each vertex. We also observe that at every edge, exactly two faces meet. We begin our construction by selecting a vertex and three non-parallel faces that meet at this vertex. Our goal is to complete the polyhedron by successively adding faces to it. It is evident that if a face shares two edges with the incomplete polyhedron, it must be part of the final structure.

To facilitate this process, we create a list $l_e$, containing the edges of the incomplete polyhedron which belong to a single face. We then search for unused faces that have at least two edges in $l_e$. This process continues until either (i) the list $l_e$ becomes empty, indicating that a closed polyhedron has been formed, or (ii) all available faces have been considered, but $l_e$ remains non-empty. In the latter case, the structure is open to infinity and is therefore discarded.

## 8. Check the origin

After finding all polyhedra created by the planes determined by the set of $n_j$ and $d_i$, we check which one of them contains the origin and return this one. If none of them contains the origin, then we return that this is an invalid set of faces.

## 9. Unnecessary faces are removed

The major drawback of using algorithm (a) is that rounded corners and edges may turn into faces which however do not represent any actual face of the original fragment. If the close enough description of the rounded body is a desired outcome one may stop here, however if the interest is in the original, fractured shape from which the scanned object evolved by weathering then step 9 must be performed. In case of algorithm (b) this step is mandatory.

In step 9, all generated faces $f_i (i=1,2,\ldots F)$ are evaluated for relevance. We check the relevance of $f_i$ by tentatively removing $f_i$ and computing the access volume $dv_i$ generated in this step and we call $f_i$ relevant if $dv_i$ is sufficiently large. We implemented this criterion by introducing the threshold $dv_0$ and for $dv_i < dv_0$ we say that the face $f_i$ is deemed irrelevant and is removed. The threshold $dv_0$, which defines the critical volume change, is set to 20%. This process is repeated until no further faces can be removed.

Figure 3 shows the volume as function of the number of faces of the polyhedron for samples 008 and 009. If the scanned object is close to a polyhedron with small number of faces (sample 009, left panel in Fig. 3) then in this process the volume remains constant until a critical face number when it starts increasing drastically. The resulting number *F* of faces is very clearly *F*=6, both from this analysis and also as concluded from the plateau length in the number of maxima plotted as a function of $\sigma$. In the right panel we see few plateaus for object 009. Visual inspection of the object also reveals this feature: the object has some small faces which may or may not be included in the final shape.

We have found that one can get rid of the parameter $dv_0$. If we consider the function *VF (*shown with the green curve in the bottom row of Fig 3. ) then the absolute minimum of this function coincides with the longest plateau on the $F(\sigma)$ histogram. We performed the analysis with this choice for the minimum and its performance is only slightly worse than method (b). It admits mostly a larger number of faces than the other two methods or the hand count, however, this almost always happens in the case of non-polyhedron like objects. This signals that this could be a desirable behavior. The supplied code has a switch to use this option.

**Summary**

Here we summarize the parameters of the algorithms (In brackets we report the values used in this presentation), recommendations for values afterwards:
Algorithm (a):

- *N: the set of number of grid points of the spherical histogram.* (values used in example: *N=500,1000,2000*), set according to the desired resolution
- *S*: the number of values for the sigma of the Gaussian kernel (*S*=50), must be set with respect to *N* to cover grid level to radius with reasonable resolution.
- $F_{max}$: The maximum number of sides considered. A reasonable cutoff speeds up the process considerably for not nice polyhedral objects. ($F_{max}=16$)
- *dn*: When making side distance histograms the limit on the 1-cosine difference of the angle between the determined normal and the face of the object (*dn* = 0.05). Should be adjusted to the resolution and desired number of faces.
- *B*: The number of automatic bins for the side distance histogram. (*B* = 10). May be increased if object's surface is rough.
- $dv_0$: Relative volume increase allowed by the side removal part ($dv_0 2$). This is the final step. which is fast and can be fitted easily. Generally there is no big difference between 0.1-0.5.

Algorithm (b):

- *N: the of number of grid points of the spherical histogram.* (*N= 2000*), set according to the desired resolution
- *S*: the number of values for the sigma of the Gaussian kernel (*S*=50), must be set with respect to *N* to cover grid level and third circle with reasonable resolution.
- $F_{max}$: The maximum number of sides considered. ($F_{max}=16$). Here it is the most important parameter, as object with this number of sides will be created and then simplified to fit.
- *dn*: When making side distance histograms the limit on the 1-cosine difference of the angle between the determined normal and the face of the object (*dn* = 0.05). Should be adjusted to the resolution and desired number of faces.

- *B*: The number of automatic bins for the side distance histogram. (*B* = 10). May be increased if object's surface is rough.
- $dv_0$: Relative volume increase allowed by the side removal part ($dv_0=0.2$). This is the final step. which is fast and can be fitted easily. Generally there is no big difference between 0.1-0.5.

**Method validation**

We have validated on a dataset used for the publication [Domokos 2020] and the data can be accessed from https://osf.io/h2ezc/. It contains 132 fragments from the Hármashatárhegy mountain in Budapest, Hungary. The composition of the fragments is dolomite. The fragments were collected from weather induced fractured areas. All fragments were scanned in three dimension and an expert determined the polyhedral structure.

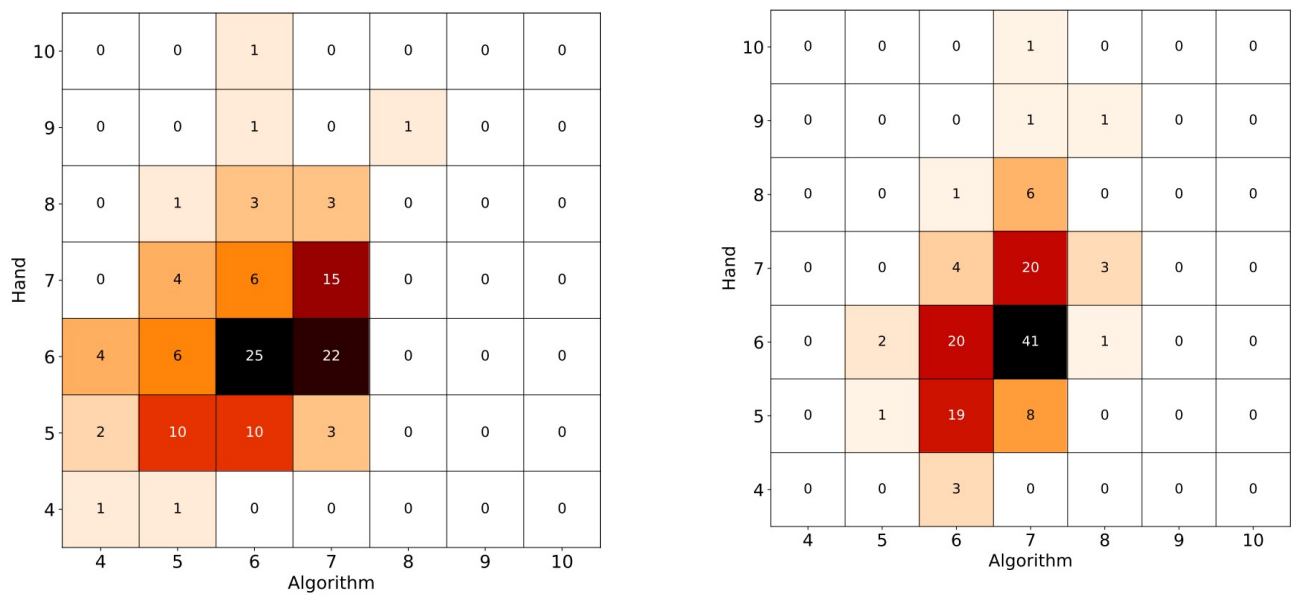

**Figure 5:** Confusion matrix between the hand measurement and the algorithm. The numbers indicate the number of pebbles in that category. Left: algorithm (a) , right: algorithm (b).

Figure 5 displays the confusion matrix of the two algorithms with respect to the hand measurement. The coloring is used to drive the eye. The value of the confusion matrix at position *i,j* shows the number of objects for which the hand measurement gave *i* the algorithm *j* sides. We have also measured two metrics, the average Hamming distance and the average Euclidean distance. The Hamming distance is 1 if both the hand and the algorithm measurements report the same number of sides, and 0 otherwise, so the larger the value the better it is. The Euclidean distance is the square root of the sum of the side difference square values also normalized by the number of samples. It is smaller for better matches. We also report the volume fraction, where the volume of the resulting polyhedron is divided by the original one. The results are summarized in the following tables.

|  | Algorithm (a) | Algorithm (b) |
|---|---|---|
| Hamming | 0.43 | 0.31 |
| Euclidean | 0.31 | 0.09 |
| Volume(average $\pm$ standard deviation) | 1.79±0.85 | 1.33±0.13 |

The table shows that although algorithm (a) is better in the Hamming distance metric, namely there are more objects where it matches exactly the experiments, when it fails it is off by a larger amount therefore the Euclidean distance is larger. The biggest difference between the two methods is in the volume of the recovered polyhedra. The biggest misses of algorithm (a) are for objects with rounded edges. In such cases algorithm (b) has superior characteristics.

**Limitations**

Our algorithm is designed to recover the polyhedral shape of scanned pebbles. It performs as intended if the object does not have rounded faces. Below, we outline some limitations:
(1) The reconstruction process is not unique, as a small face may result from the original fragmentation or, alternatively, from subsequent weathering. The algorithm reflects this natural ambiguity, however, it also provides certain parameters to control this, namely its two versions and the parameters $dv_0$ and $F_{max}$.
(2) We have compared the results of the algorithms to human measurements, while we also acknowledge that the latter can also be problematic in some cases. Figure 7 displays two examples where the two algorithms agree with each other but not with the human expert. Our perspective is that, in these cases, the algorithm outperformed the human expert.
(3) In some cases, algorithm (a) did not return a fit at all; we list these objects in Fig. 8. Obviously, these objects are very rounded and although algorithm (b) returns a fit for them their usefulness is questionable. Therefore we conclude that if algorithm (a) fails to return a polyhedral fit, we may consider the polyhedral approximation of the object is not feasible.

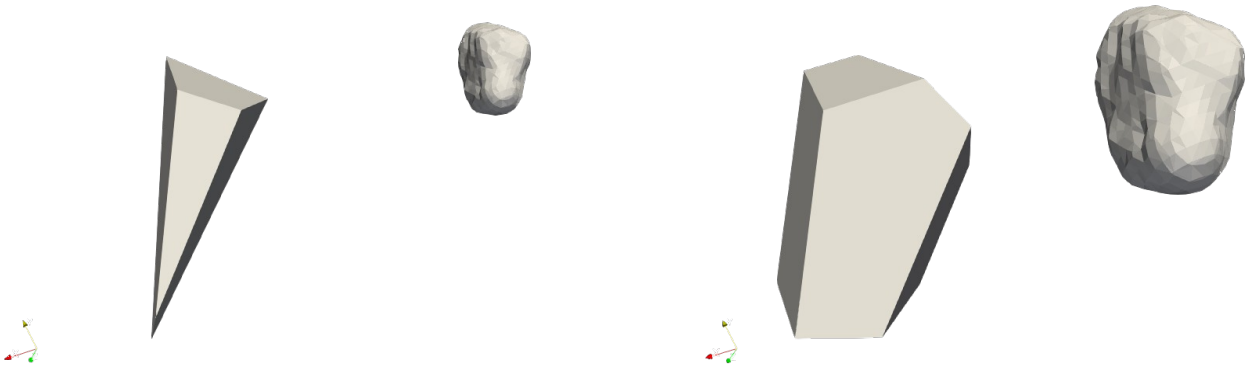

**Figure 6:** Results of the two methods for object 40. The left panel shows the results of algorithm (a) with 4 sides, the right with algorithm (b) with 7 faces.

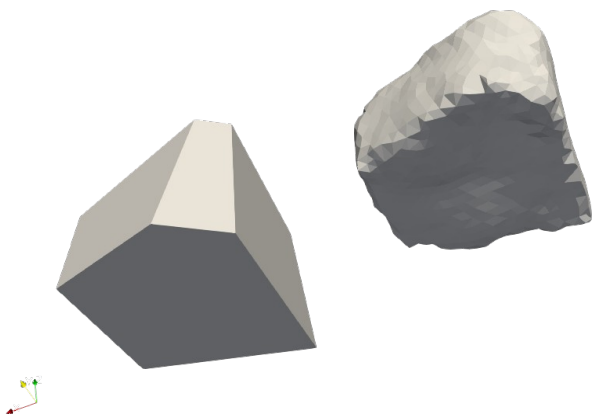

**Figure 7:** Result for the object 4 for both algorithms with 7 faces. The face in the front was not identified by the expert.

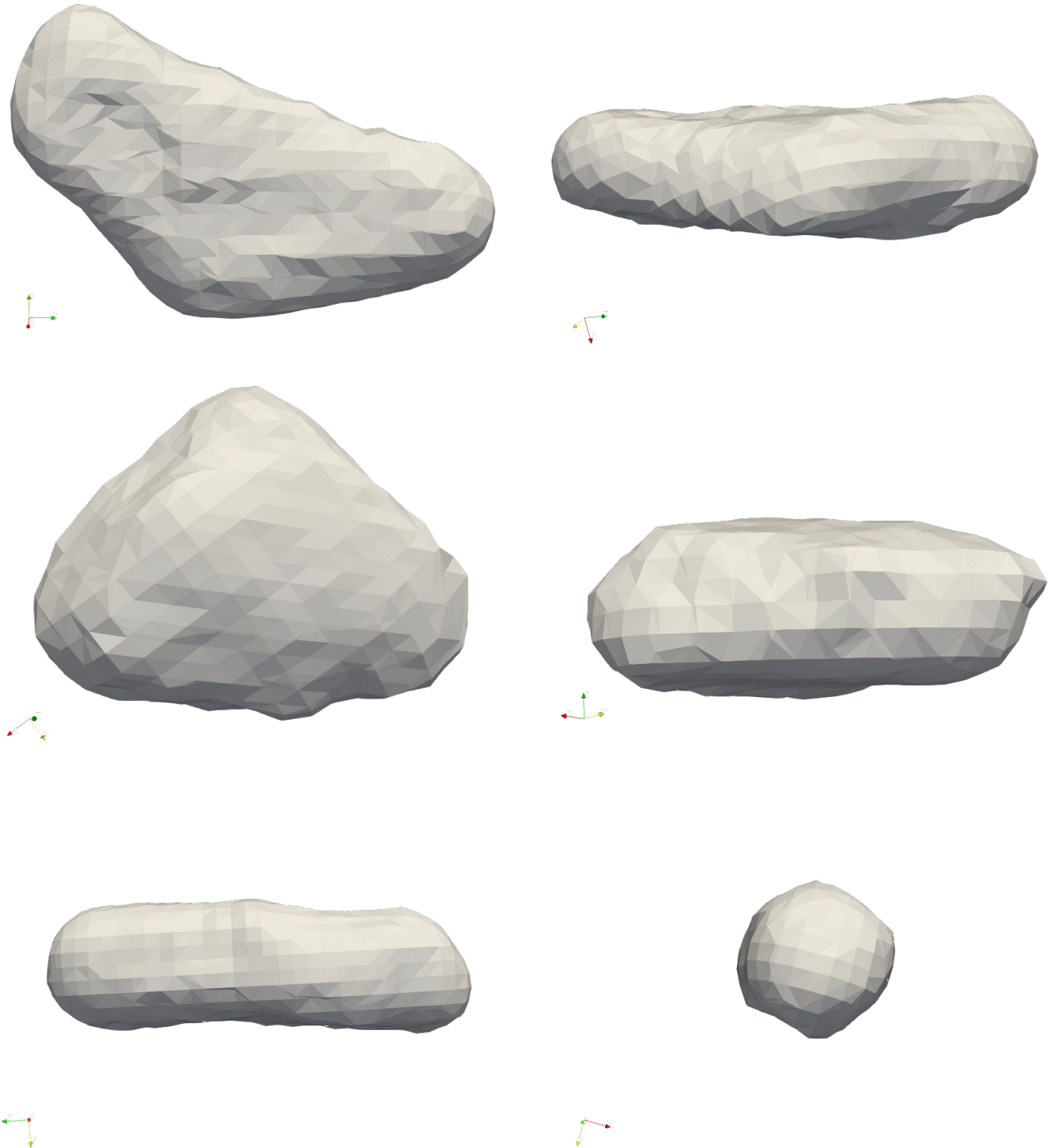

**Figure 8:** Front and side view of objects which fail with algorithm (a), for objects 54, 58, 86 from top to bottom respectively.

## Ethics statements
None

## CRediT author statement
**János Török**: Conceptualization, methodology, code development, writing and editing, **Gábor Domokos**: Conceptualization, methodology, writing and editing

## Acknowledgments
Support from Advanced Grant 149429 from the Hungarian National Research, Development and Innovation Office and Grant TKP2021-NVA-02 of the National Research, Development and Innovation Fund of Hungary to G.D is gratefully acknowledged, further support was provided by Thematic Area Excellence Program of the Ministry for Culture and Innovation from NRDI Fund (Grant No.: TKP2021-NVA-02).

## Declaration of interests

☒ The authors declare that they have no known competing financial interests or personal relationships that could have appeared to influence the work reported in this paper.

☐ The authors declare the following financial interests/personal relationships which may be considered as potential competing interests: